\documentclass[epj,final]{svjour}

\usepackage{latexsym}
\usepackage{url}
\usepackage{amsfonts}

\RequirePackage{graphicx}

\begin{document}
\title{On the common nature of dark matter and dark energy: galaxy groups}
\author{V.G. Gurzadyan
}                     
%
%
\institute{Center for Cosmology and Astrophysics, Alikhanian National Laboratory and Yerevan State University, Yerevan, Armenia}
\date{Received: date / Revised version: date}
%

\abstract{
It is shown that the cosmological constant links the roots both of General Relativity and Newtonian gravity via the general function satisfying Newton's theorem according to which the gravitating sphere acts as a point mass situated in its center. The quantitative evidence for this link is given via the correspondence of the current value of the cosmological constant with the value of the cosmological term in the modified Newtonian gravity to explain the dark matter in the galaxies. This approach reveals:  (a) the nature of dark matter as gravity's signature, (b) the common nature of dark matter and of cosmological constant (dark energy), (c) that the dark matter as a consequence of repulsive gravity increasing with the distance ensures the observed higher mass-to-luminosity M/L ratio while moving from the scales of galaxies to galaxy clusters. The galactic halos via non-force-free interaction due to the repulsive $\Lambda$-term of gravity determine the features of galactic disks as is supported by observations. The data of galaxy groups of the Hercules--–Bootes region are also shown to support the $\Lambda$-gravity nature of the dark matter i.e. the value of the cosmological constant is derived not from cosmological data. Among the consequences of such modified General Relativity is the natural link to AdS/CFT correspondence.        
} 

%
\maketitle
\section{Introduction}

The observational indications for the dark matter and dark energy have stipulated the active development of variety of models, including those of modifications of Newton's gravity and of General Relativity (GR).
Various principles are taken as bases in those models, such as scalar-tensor, f(R) theories, MOND (see  \cite{Sand,Clif,Bert,Bah} and refs therein)  with different motivations and with the natural aim to satisfy the observational data or to suggest verification tests for future observations.

Below we reconsider a principle which is in the very roots of Newton's gravity, i.e. the theorem that a sphere is acting gravitationally as a point mass situated in its center. That theorem had enabled Newton to attribute the gravitational law $r^{-2}$ to the motion of planets which are definitely extended spheres and are not point masses. Later the Newton's gravity became a key element GR, acting as its weak field limit and enabling to correspond the predicted effects with the observed ones.  That concerns, however, the Einstein equations without the cosmological term and Einstein's motivation \cite{E} for introducing the cosmological constant was the static Universe. In the approach below we show that the cosmological term appears in Einstein's equations from the above mentioned theorem proved in \cite{N}.  

As shown in \cite{G}, the most general function satisfying that theorem, besides the usual $r^{-2}$ term, contains also another term with a cosmological constant. 

Taking that modified Newton's law with a cosmological constant as the weak field limit of GR one arrives to a modification of GR containing the cosmological term naturally! We then show that this approach both to Newton's gravity theory and to GR enables one to link two observational facts, i.e. the dark matter in the galaxies and the cosmological constant. The dark matter then appears as an observational contribution of repulsive gravity at large scales, i.e. in galactic halos and clusters of galaxies.

\section{Newton's theorem and General Relativity}

As shown in \cite{G} the most general function for the force satisfying Newton's theorem i.e. the condition for the sphere to attract as a point of the sphere's mass and situated in its center has the form 
\begin{equation}
f(r)= Ar^{-2} + \Lambda r,
\end{equation}   
as the solution of equation
\begin{equation}
\frac{r^2}{2}f''(r) + r f'(r)-f(r)=0,
\end{equation}
where $A$ and $\Lambda$ are constants.

The second term corresponds the cosmological constant term if one turns to the Newtonian form of the Friedmann cosmological equation \cite{G}. 
Eq. (1), however, within the context of the shell theorem defines a force-free field only in the center of a shell, but preserves the O(3) symmetry.  

Turning to the GR, instead of the usual Newtonian limit for its weak field approximation \cite{W} 
now metric tensor components $g_{00}$ and $g_{rr}$ will be modified and one will have the metric for the point mass as 
\begin{equation}
ds^2 = (1 - 2A r^{-1} - \Lambda r^2/3)c^2 dt^2 + (1 - 2A r^{-1} - \Lambda r^2/3)^{-1}dr^2 + r^2 d\Omega^2.
 \end{equation}   
The principal fact here is in the following. The Einstein equations without cosmological term are considered to have the usual Newtonian limit in weak-field approximation, while the Einstein equations with cosmological term will formally violate that limit, e.g. according to Weinberg  "{\it ..$\Lambda$ must be very small so as not to interfere with the successes of Newton's theory of gravitation.}" (Chapter 7.1, \cite{W}). We now see that via Eq.(1) that violation is removed. 
Eq.(3) is the covariant metric having its weak-field limit Eq.(1) \cite{No}.

Instead, we see that Eq.(1) ensures the weak-field limit for Einstein equations with cosmological constant 
\begin{equation}
G_{\mu\nu} + \Lambda g_{\mu\nu}= \kappa T_{\mu\nu}.   
\end{equation}  

The adoption of the Newton's "sphere = point mass" theorem and hence of the Eq. (1) will readily lead to renormalization in various predictions of GR.  Although such modification of GR coincides with its original form of Einstein's equations,  there is drastic difference in the motivations.

Here the starting point is the Newton's gravity and his sphere-point theorem and $\Lambda$ is appearing in Einstein's equations readily and not as an extra term added by hand to fulfill a static universe concept.  Namely, if Newton might have found the general function Eq. (1), then Einstein initially would have written GR equations with $\Lambda$ i.e. with the link of O(3) and the Lorentz group. 

Of course, the $\Lambda$-term in Eq.(3) has been considered previously (Schwarzschild -- de Sitter metric), however, the Newton theorem's approach described here, as we will see below leads to insights on the common nature of the dark matter and dark energy (cosmological constant). 

\section{The Eq.(1), the cosmological constant and the dark matter}   

If the Einstein equations with the cosmological constant have the Newtonian limit Eq.(1), then one will have a link e.g. to the two currently adopted principal observational facts on the dark energy and the dark matter. Indeed, while the cosmological constant in the Einstein equations is considered to describe the acceleration of the Universe, the Newtonian potential and its modifications are attributed e.g. to the observational indications for the dark matter in the galaxies.

The value of the cosmological constant is deduced in several ways, the one indicated by the Planck data \cite{P} is
\begin{equation}
\Lambda \simeq 1.1\,\, 10^{-52} m^{-2}.
\end{equation}  
Regarding the dark matter in the galaxies, it is shown that the parameters of the halos determine the late disk and early spheroidal structures of the galaxies \cite{Kr}. Then, for virialized  (for "oscillator" term $\propto R^k$ with $k=2$, see Eq. (10.7) in \cite{LL}) structures we have
\begin{equation}   
\Lambda=\frac{3\sigma^2}{2 c^2 R^2}\simeq 3\,\, 10^{-52} (\frac{\sigma}{50\, km s^{-1}})^2(\frac{R}{300 \,kpc})^{-2} m^{-2},
\end{equation} 
normalized to the velocity dispersion at a given radius of halo. 

Numerically $\Lambda$s in (5) and (6) are close and this fact can be interpreted as follows. The positive cosmological constant corresponds to the accelerating Universe and hence to negative pressure and to repulsion as evidence of vacuum energy \cite{Z}. The crucial point regarding the dark matter is that Eq.(1) defines non-force-free field within the sphere except its center, increasing from center, thus mimicking increase of the central mass. Thus, within this interpretation the dark matter is a gravitating mass with force (1) and revealing its repulsion at large scales e.g. in galactic halos.  The effective increase of the central attracting mass will support the effect of "flat rotation curves", although, obviously, the description of given observational rotation curves will need extensive modeling and numerical simulations for the input parameters of the disk and halo. For the analysis of the dark matter problem within $f(R)$ theories see \cite{Cap}.  

\section{Galaxy groups and Eq.(6)}

We now test this approach on $\Lambda$-nature of dark matter using the data by Karachentsev et al \cite{Kar} for a sample of 17 galaxy groups of Hercules–--Bootes region which (the data) include the $\it rms$ galactic velocities $\sigma$ and the harmonic average radii $R_h$ of the groups.  Then, using Eq.(6) we obtain the $\Lambda$ as presented in the Table 1; the galaxy groups are denoted by the name of the brightest galaxy (first column). 

\begin{center}
 {\bf Table 1.} $\Lambda$ obtained using Eq.(6) for galaxy groups of the Hercules---Bootes region.
 {
 \renewcommand{\baselinestretch}{1.2}
 \renewcommand{\tabcolsep}{3.5mm}
 \small
 \begin{tabular}{ | l | r | r | r | }
 \hline
    Galaxy group    &  $\sigma (km/s^{-1})$ & $R_h(kpc)$ & $\Lambda(m^{-2})$\\ \hline		
		\hline
    NGC4736  &   50 & 338 &  3.84E-52 \\ \hline
    NGC4866  & 	 58 & 168 &  2.09E-51 \\ \hline
    NGC5005  & 	114 & 224 &  4.55E-51 \\ \hline
    NGC5117  & 	 27 & 424 &  7.12E-53 \\ \hline
    NGC5353  & 	195 & 455 &  3.23E-51 \\ \hline
    NGC5375  & 	 47 &  66 &  8.91E-51 \\ \hline
    NGC5582  & 	106 &  93 &  2.28E-50 \\ \hline
    NGC5600  & 	 81 & 275 &  1.52E-51 \\ \hline
    UGC9389  & 	 45 & 204 &  8.55E-52 \\ \hline
    PGC55227 & 	 14 &  17 &  1.19E-50 \\ \hline
    NGC5961  & 	 63 &  86 &  9.43E-51 \\ \hline
    NGC5962  & 	 97 &  60 &  4.59E-50 \\ \hline
    NGC5970  & 	 92 & 141 &  7.48E-51 \\ \hline
    UGC10043 & 	 67 &  65 &  1.87E-50 \\ \hline
    NGC6181  & 	 53 & 196 &  1.28E-51 \\ \hline
    UGC10445 & 	 23 & 230 &  1.76E-52 \\ \hline
    NGC6574  & 	 15 &  70 &  8.07E-52 \\ \hline
		\hline
    Average  &      &     &  8.24E-51 \\ \hline
    St.deviation	   &      &     &  1.15E-50 \\ \hline
		\hline
 \end{tabular}
 }
 \end{center}

The correspondence of values of $\Lambda$ in Table 1 to those in Eqs.(5) and (6) is visible.

\section{Conclusions}

We demonstrated a natural way for the appearance of the cosmological constant in Einstein equations. This follows while adopting for the weak field approximation of General Relativity the Eq.(1) as the general function satisfying the Newton's theorem of 1687, that the gravitating sphere acts as a point mass located in its center. This drastically differs from Einstein's original motivation for the introduction of the cosmological constant.

This approach enables to draw the following conclusions:   

(a) the nature of the dark matter in galaxies is in the gravity;

(b) the dark energy (cosmological constant) and the dark matter are of the common nature;  

(c) the dark matter is the signature of repulsive gravity of the ordinary matter determined by $\Lambda$ constant and dominating at large scales, i.e. larger than of galactic halos.

Thus, the key constituents of the dark universe are naturally linked here. The repulsive gravity nature of the dark matter is responsible for the observed increase of the mass-to-luminosity M/L ratio while moving from the scales of galaxies to those of galaxy clusters. The non-force-free shell (galactic halo) determines the internal structures of galaxies (disks), as indicate the observations.
The observational data of a sample of galaxy groups \cite{Kar}, i.e. systems containing 3 or more galaxies, are shown to confirm Eq.(6) on the $\Lambda$-nature of the dark matter.

Obviously, far more consequences of this modified GR, i.e. with weak-field of Eq.(1), are of interest regarding the experimental and theoretical aspects.  The current experimental tests for GR, including the recent 5\% accuracy for the Lense-Thirring effect obtained via the LARES satellite \cite{Lares}, are obviously far from detecting the potential contribution of the $\Lambda$-term in GR. However, the observations of black holes and pulsars including the detections of gravitational waves via LIGO \cite{LIGO}, on the one hand, and of galactic halos (e.g. \cite{G2}), dynamics of galaxy groups and clusters for weak-field gravity, on the other hand, can provide efficient tests. Then, the cosmological constant determines not only expansion of the Universe but also the weak-field gravity and even possibly is linked to the arrow of time \cite{AG}. The modified GR reveals also the obvious link to the AdS/CFT correspondence, since instead of Poincare group now one has a SO(2,3) group.  Further consequences of this approach are in \cite{GS1,GS2}.

\end{document}